\documentclass[aps,prd,showpacs,floatfix]{revtex4}
\usepackage{amsmath,amsfonts}
\usepackage[dvips]{graphicx}

\newcommand{\ud}{\mathrm{d}}
\newcommand{\tr}{\mathrm{tr}}

\begin{document}

\title{Extended Quintessence with non-minimally coupled phantom scalar field}
\author{Orest Hrycyna}
\email{hrycyna@kul.lublin.pl}
\affiliation{Department of Theoretical Physics, Faculty of Philosophy, 
The John Paul II Catholic University of Lublin, Al. Rac{\l}awickie 14, 
20-950 Lublin, Poland}
\author{Marek Szyd{\l}owski}
\email{uoszydlo@cyf-kr.edu.pl}
\affiliation{Astronomical Observatory, Jagiellonian University,
Orla 171, 30-244 Krak{\'o}w, Poland}
\affiliation{Mark Kac Complex Systems Research Centre, Jagiellonian University,
Reymonta 4, 30-059 Krak{\'o}w, Poland}


\begin{abstract}
We investigate evolutional paths of an extended quintessence with a non-minimally
coupled phantom scalar field $\psi$ to the Ricci curvature. The dynamical system
methods are used to investigate typical regimes of dynamics at the late time. We
demonstrate that there are two generic types of evolutional scenarios which 
approach the attractor (a focus or a node type critical point) in the phase space: the 
quasi-oscillatory and monotonic trajectories approach to the attractor which
represents the FRW model with the cosmological constant. We demonstrate that 
dynamical system admits invariant two-dimensional submanifold and discussion 
that which cosmological scenario is realized depends on behavior of the system on 
the phase plane $(\psi, \psi')$. We formulate simple conditions on the value of
coupling constant $\xi$ for which trajectories tend to the focus in the phase plane and hence damping 
oscillations around the mysterious value $w=-1$. We describe this condition in 
terms of slow-roll parameters calculated at the critical point. We discover 
that the generic trajectories in the focus-attractor scenario come from the 
unstable node. It is also investigated the exact form of the parametrization of 
the equation of state parameter $w(z)$ (directly determined from dynamics) 
which assumes a different form for both scenarios.
\end{abstract}


\maketitle

\section{Introduction}

The non-minimal coupling of the type $\frac{1}{2}\xi R \psi^{2}$ in the 
Lagrangian, where $R$ is the Ricci scalar and $\xi$ -- the parameter of 
non-minimal coupling was independently introduced by Chernikov and Tagirov 
\cite{Chernikov:1968zm} and Callan et al. \cite{Callan:1970ze}. It was pointed 
out that we should consider $\xi \ne 0$ and different arguments were given. The 
nonzero $\xi$ arises from quantum corrections \cite{Birrell:1984ix} and it is 
required by the renormalization \cite{Callan:1970ze}. Faraoni 
\cite{Faraoni:2000gx} argued that the non-minimal coupling of the inflaton 
field is unavoidable and in his opinion the paradigm of standard inflation 
should be generalized by including the coupling term in the Lagrangian. If the 
parameter $\xi$ assumes a value different from $1/6$ (conformal coupling case) 
then the equation of motion for the scalar field (i.e., the Klein-Gordon
equation) is conformally invariant if $V(\psi)=0$ or $V(\psi)\propto\psi^{4}$
\cite{Wald:1984rg}. While there are strong arguments that seems to favor the
choice of $\xi=1/6$, under the assumption of conformal invariance of the 
Klein-Gordon equation and that $\psi$ does not violate the equivalence principle
\cite{Faraoni:1996rf,Sonego:1993fw} (contrary to Lightman et al. 
\cite[p. 85]{Lightman:1975}). Moreover the value of $\xi=1/6$ cannot be treated
as a correct coupling for various scalar field particles.
Adopting a different point of view \cite{Faraoni:1997fn} it is desirable to
study different constraints on the coupling constant. The strength of the 
non-minimal coupling is constrained to be $\xi > -7.0*10^{-3}$ for negative
$\xi$ in the case of a quadratic potential function \cite{Tsujikawa:2004my}. 

Recent astronomical observations (like SNIa, the CMB and other
\cite{Riess:1998cb, Astier:2005qq, Davis:2007na, Spergel:2006hy}) indicate that
current universe is almost flat and undergoing accelerated phase of expansion.
While the cosmological constant term introduced to the equations of general relativity 
offers the possibility of explanation of acceleration, its value is 
extraordinary small compared with an inferred vacuum energy. To solve this 
problem it is natural to consider a source of this acceleration in the form of 
a scalar field (quintessence idea \cite{Zlatev:1998tr, Carroll:1998zi}). In 
the simplest case minimally coupled to gravity scalar fields with a potential 
$V(\psi)$ are considered. In this case the dynamical effect of a scalar field 
is equivalent to a perfect fluid with the energy density and the pressure 
given by $\rho_{\psi}=\frac{1}{2}\dot{\psi}^{2} + V(\psi)$ and
$p_{\psi}=\frac{1}{2}\dot{\psi}^2 - V(\psi)$, respectively. In the canonical
quintessence scenario the kinetic energy is $k=\dot{\psi}^{2}/2$ and the true 
cosmological constant is zero and the potential (tracking potential) is 
slow rolling down \cite{Bludman:2003rv}. 

This quintessence field generically has good-or-bad attractor properties
\cite{Ratra:1987rm, Wetterich:1987fm}, i.e., it is sensitive with respect to
choice of the initial conditions. The tracking quintessence was invoked to use this
attractor property to explain the smallness of the cosmological constant without
fine tuning of the potential or initial conditions.
The idea of quintessence was extended to the case of non-minimally coupled
scalar field \cite{Faraoni:2000wk, Uzan:1999ch, Amendola:1999qq}. The main
motivation of this extension was to explain the
observational data which favor the negative equation of state parameter $w=p/\rho <
-1$ \cite{Choudhury:2003tj}. If we consider the canonical scalar field in the
background of Einstein gravity then we cannot explain the range of value $w <
-1$. If we assume the flat universe then $w<-1$ implies superacceleration as a
opposed to acceleration $\ddot{a}/a = \dot{H}+H^{2}>0$, where $a$ is the scale
factor and $H=(\ln(a))\dot{}$ is the Hubble parameter. There are two natural
possibilities of extension of the theory: 1) to allow the scalar field $\psi$ to
couple non-minimally to the Ricci scalar \cite{Faraoni:2006ik}, and 2) including
of the phantom scalar fields, i.e., the scalar field is formally allowed to have a
negative kinetic energy and $\dot{\psi}^2$ is replaced by $(-)\dot{\psi}^{2}$
\cite{Caldwell:1999ew, Dabrowski:2003jm}.
This type of modification has a justification from string, M theory and in
supergravity. Note that such possibilities can explain the breaking $w<-1$
barrier. Both phantom and non-minimally coupled scalar field models can be 
regarded as a scalar tensor theories of gravity. Such models are called 
extended quintessence \cite{Faraoni:2006ik, Faraoni:2004, Fujii:2003}.
Recently Fuzfa and Alimi suggested a promising unified description of both dark
matter and dark energy. It relies on a violation of the weak equivalence
principle on cosmological scales by dark matter. It is interesting that amplitude
of this violation depends on the relative concentration of baryons
\cite{Fuzfa:2007sv}.

The main aim of this paper is to investigate the long time evolution of 
extended quintessence models. We give the complete qualitative classification 
of solutions according to critical point approach scenarios (the monotonic 
approach to a critical point or the quasi-oscillatory regime of approaching 
and crossing of the $w=-1$ barrier). We use the dynamical system methods in 
investigation of all evolutional paths for all admissible initial conditions. 
These methods were previously used (see for example 
\cite{Gunzig:2000kk, Gunzig:2000ce}) in the context of the FRW model with
non-minimally coupled scalar fields and superacceleration. We are interested in
typical evolutional scenarios of the phantom non-minimally coupled scalar field
at a late time near the $w=-1$ state. We demonstrate that there are two
different scenarios to achieve this state. The new scenario of the route to the 
cosmological constant is through the damping oscillations of the scalar field
asymptotically going to the constant value. In this scenario there is an 
infinite number of crossing the phantom divide line $w=-1$. 
In our previous paper \cite{Hrycyna:2007mq} we study how deSitter state can be
achieved in cosmology with conformally coupled to gravity phantom scalar field. 
The dynamics of the phantom scalar field conformally coupled to gravity with
quadratic potential function depends on the value of squared mass of the field.
The dynamics is regular for $m^{2}>0$, in opposite case with $m^{2}<0$ the dynamics becomes
chaotic and fractals structure appears in the phase space
\cite{Szydlowski:2006qn}.
In the present 
paper we generalize result of \cite{Hrycyna:2007mq} by consideration $\xi$ as a
free parameter which
should be constrained by observational data. In investigation of dynamics with the
help of dynamical system methods this parameter plays the role of bifurcation
parameter and we found some critical value of this parameter for which
topological structure of phase space changes.
We show that damping oscillations around $w=-1$ value
type of evolution can be described by a critical point of a focus type on a 
$2$-dimensional invariant submanifold $(\psi,\psi')$ where 
$'\equiv \ud/\ud \ln(a)$. We will demonstrate that the late time evolutionary 
scenario crucially depends on the value of the non-minimal coupling constant.

\section{FRW model with the non-minimally coupled phantom scalar field}

We assume the flat model with the FRW geometry, i.e., the line element has the 
form
\begin{equation}
\ud s^{2} = -\ud t^{2} + a^{2}(t)[\ud r^{2} + r^{2}(\ud\theta^{2}
+ \sin^{2}{\theta}\ud\varphi^{2})],
\label{eq:1} 
\end{equation}
where $0 \leq \varphi \leq 2\pi$, $0 \leq \theta \leq \pi$ and $0 \leq r \leq
\infty$ are comoving coordinates, $t$ stands for the cosmological time.
It is also assumed that a source of gravity is the phantom scalar field
$\psi$ with an arbitrary coupling constant $\xi$. The dynamics is governed
by the action
\begin{equation}
S=\frac{1}{2}\int \ud^{4}x \sqrt{-g}\Big(m_{p}^{2}R +
(g^{\mu\nu}\psi_{\mu}\psi_{\nu} + \xi R\psi^{2} - 2 U(\psi))\Big)
\label{eq:2}
\end{equation}
where $m_{p}^{2}=(8\pi G)^{-1}$; for simplicity and without lost of generality
we assume $4\pi G/3=1$ and $U(\psi)$ is a scalar field potential.
After dropping the full derivatives with respect to time, rescaling the time 
variable to the conformal time $ \ud\eta = \ud t/a$ we obtain the energy 
conservation condition
\begin{equation}
\mathcal{E} = \frac{1}{2}\dot{a}^{2} + \frac{1}{2}a^{2}\dot{\psi}^{2} +
3\xi\dot{a}^{2}\psi^{2}+6\xi\dot{a}\dot{\psi}a\psi - a^{4}U(\psi)
-\rho_{m,0}a=\rho_{r,0}
\label{eq:3}
\end{equation}
where $\rho_{r,0}$ and $\rho_{m,0}$ are the constants corresponding to the
radiation and the matter in the model, respectively, and dot denotes
differentiation with respect to conformal time.
We can also express this condition as a
\begin{equation}
\frac{1}{2}H^{2} = \rho_{\psi} + \rho_{r} + \rho_{m}
\label{eq:4}
\end{equation}
where $H$ is the Hubble function, $\rho_{\psi}$ is the energy density of a 
phantom scalar field, $\rho_{r}\propto a^{-4}$ is the energy density of 
radiation and $\rho_{m}\propto a^{-3}$ is the density of matter.

The Euler-Lagrange equation reduces to
\begin{equation}
\left\{ \begin{array}{l}
a^{2}\ddot{\psi} + 6\xi\ddot{a}a\psi = -2\dot{a}\dot{\psi}a + a^{4}U(\psi), \\
\ddot{a}(1+6\xi\psi^{2}) + 6\xi\ddot{\psi}a\psi = a\dot{\psi}^{2}(1-6\xi) -
12\xi\dot{a}\dot{\psi}\psi + 4a^{3}U(\psi) + \rho_{m,0}.
\end{array} \right.
\label{eq:5}
\end{equation}
It is easy to check that for the conformal coupling $\xi=1/6$, quadratic potential
function, the rescaled scalar field variable $\psi=\phi/a$ and $\rho_{m,0}=0$ this
system reduces to well known one.

It would be useful to change
\begin{align}
\dot{\psi} & =\frac{\dot{a}}{a}\psi', \nonumber \\
\ddot{\psi} & =\frac{\ddot{a}}{a}\psi'+\frac{\dot{a}^{2}}{a^{2}}(\psi''-\psi'),
\nonumber
\end{align}
where a dot denotes differentiation with respect to the conformal time and a 
prime with respect to a natural logarithm of the scale factor. Equations of 
motion (5) are in the form
\begin{equation}
\left\{ \begin{array}{l}
\frac{\ddot{a}}{a^{3}}(\psi'+6\xi\psi)+\frac{\dot{a}^{2}}{a^{4}}(\psi''+\psi') =
U'(\psi),\\
\frac{\ddot{a}}{a^{3}}(1+6\xi\psi(\psi'+\psi)) +
6\xi\psi\frac{\dot{a}^{2}}{a^{4}}(\psi''+\psi') =
\frac{\dot{a}^{2}}{a^{4}}\psi'^{2}(1-6\xi) + 4U(\psi) +\rho_{m},
	\end{array} \right.
\label{eq:6}
\end{equation}
where
\begin{equation}
\frac{\dot{a}^{2}}{a^{4}}=2\frac{U(\psi)+\rho_{r}+\rho_{m}}
{1+(1-6\xi)\psi'^{2}+6\xi(\psi'+\psi)^{2}}.
\label{eq:7}
\end{equation}
After eliminating the scale factor and it derivatives we obtain
\begin{align}
&(\psi'' + \psi')\left[1+6\xi(1-6\xi)\psi^{2}\right] +
\psi'^{2}(1-6\xi)(\psi'+6\xi\psi) + \nonumber \\ 
&+ \frac{1}{2}
\frac{1+(1-6\xi)\psi'^{2}+6\xi(\psi'+\psi)^{2}}{U(\psi)+\rho_{r}+\rho_{m}}
\bigg[ (4U(\psi)+\rho_{m})(\psi'+6\xi\psi) -
U'(\psi)(1+6\xi\psi(\psi'+\psi))\bigg] = 0,
\label{eq:8}
\end{align}
where a prime denotes the differentiation with respect to a natural 
logarithm of the scale factor. 

The dynamical systems methods offer a possibility of analyzing all evolutional 
paths for all admissible initial conditions in a phase space. They are 
represented by the trajectories in the phase space. The structure of the phase 
space is organized by the trajectories and limit sets. The critical points of 
the system are such points in the phase space for which right-hand sides of 
the dynamical system vanishes at these points. If the system is non-linear we 
can linearize it at these critical points and the famous Hartman-Grobman 
theorem guarantees that linearized system is a good approximation of the 
original dynamical system in the vicinity of this point. The type of a critical 
point is also characterized by a linearization matrix. It can be determined 
from the characteristic equation $\det|A-\lambda\mathbf{1}|=0$, where $A$ is 
the linearization matrix. The critical point is the global attractor if all 
real parts of the eigenvalues are negative, i.e., 
$\mathrm{Re}\lambda_{i}<0$ $\forall_i$.

Introducing the new variable $y=\psi'$ we can represent equation (\ref{eq:8}) 
as the autonomous dynamical system
\begin{align}
\psi' &= y \nonumber \\
y' &= -y- y^{2}(y+6\xi\psi)\frac{1-6\xi}{1+6\xi\psi^{2}(1-6\xi)} - \nonumber \\
&- \frac{1}{2}
\frac{1+(1-6\xi)y^{2}+6\xi(y+\psi)^{2}}{1+6\xi\psi^{2}(1-6\xi)}
\bigg[ \frac{(4U(\psi)+\rho_{m})(y+6\xi\psi) -
U'(\psi)(1+6\xi\psi(y+\psi))}{U(\psi)+\rho_{r}+\rho_{m}}\bigg] \\ \nonumber 
\rho_{r}' &= -4\rho_{r} \\ \nonumber
\rho_{m}' &= -3\rho_{m}, \nonumber 
\label{eq:9}
\end{align}
where we have $\rho_{m}=\rho_{m,0}a^{-3}$ and $\rho_{r}=\rho_{r,0}a^{-4}$.

The equation of state can be directly determined from the dynamics
\begin{align}
w_{\text{eff}} = 2 \Big\{ & -\xi(\psi'+\psi)^{2} + \frac{1}{H^{2}}\Big(2\xi\psi
U'(\psi)-U(\psi)+\frac{1}{3}\rho_{r}\Big) + \nonumber \\
 & + 2\frac{1-6\xi}{1+6\xi\psi^{2}(1-6\xi)}
\Big[ -\frac{1}{4}\psi'^{2}\Big(1+2\xi\psi^{2}(1-6\xi)\Big) +
\xi\frac{\psi^{2}}{H^{2}}\Big(\rho_{m}+4U(\psi)-6\xi\psi U'(\psi)\Big)\Big]\Big\}.
\end{align}
where
$$
H^{2}=2\frac{U(\psi)+\rho_{r}+\rho_{m}}{1+(1-6\xi)\psi'^{2}+6\xi(\psi'+\psi)^{2}}.
$$
is the counterpart of the Friedmann first integral.

On the invariant submanifold we change the ``time'' variable 
$$
\varepsilon\frac{\ud}{\ud \sigma}=U(\psi)(1+6\xi\psi^{2}(1-6\xi))\frac{\ud}{\ud \ln(a)}
$$
where $\varepsilon=1$ when $U(\psi)(1+6\xi\psi^{2}(1-6\xi))>0$ and
$\varepsilon=-1$ when $U(\psi)(1+6\xi\psi^{2}(1-6\xi))<0$. 
The dynamical system is in the form
\begin{align}
\varepsilon\dot{\psi}=& y U(\psi)(1+6\xi\psi^{2}(1-6\xi)),\\ \nonumber
\varepsilon\dot{y}=&-y U(\psi)(1+6\xi\psi^{2}(1-6\xi)) - (1-6\xi)y^{2}(y+6\xi\psi)U(\psi)
- \\ \nonumber
&- \frac{1}{2}(1+(1-6\xi)y^{2}+6\xi(y+\psi)^{2})(4U(\psi)(y+6\xi\psi) -
U'(\psi)(1+6\xi\psi(y+\psi))).\nonumber
\end{align}

In the generic case critical points of a dynamical system are located: \\
1) $y_{0}=0$ and $(1+6\xi\psi_{0}^{2})(24\xi\psi_{0}U(\psi_{0}) -
U'(\psi_{0})(1+6\xi\psi_{0}^{2}))=0$; \\
2) $\psi_{0}^{2}=\frac{1}{6\xi(6\xi-1)}$ for $\xi>1/6$ or $\xi<0$ and $y_{0}$ 
is a solution of the cubic equation
$$
A(\xi)y^{3}+B(\xi)y^{2}+C(\xi)y+D(\xi)=0
$$
where for a given form of the potential function, $A(\xi)$, $B(\xi)$, $C(\xi)$,
$D(\xi)$ are functions of the coupling constant $\xi$ only.

For our further investigations the critical point $(1)$ is especially
interesting. At this point the trace and the determinant of the linearization 
matrix are
\begin{equation}
\tr{A}= -U(\psi_{0})(1+6\xi\psi_{0}^{2}(1-6\xi)) -
\frac{1}{2}(1+6\xi\psi_{0}^{2}) (4U(\psi_{0})-6\xi\psi_{0}U'(\psi_{0}))
\end{equation}

\begin{equation}
\det{A}=-\frac{1}{2}(1+6\xi\psi_{0}^{2})(12\xi\psi_{0}U'(\psi_{0})+ 24\xi
U(\psi_{0})-U''(\psi_{0})(1+6\xi\psi_{0}^{2}))
U(\psi_{0})(1+6\xi\psi_{0}^{2}(1-6\xi))
\end{equation}
The type of behavior around this point crucially depends on the sign of 
$\Delta = (\tr{A})^{2}-4\det{A}$. We are interested in the case $\xi>0$. Note
that at this critical point
$$
L=\frac{U'(\psi_{0})}{U(\psi_{0})}=\frac{24\xi\psi_{0}}{1+6\xi\psi_{0}^{2}}
$$
we obtain that
\begin{equation}
\Delta = (U(\psi_{0}))^{2}
\bigg((1+6\xi\psi_{0}^{2}(1-6\xi))\Big(9(1+6\xi\psi_{0}^{2}(1-6\xi))-
48(18\xi^{2}\psi_{0}^{2}+\xi-\Gamma 24\xi^{2}\psi_{0}^{2})\Big) \bigg)
\end{equation}
The general condition for $\Delta\le0$ is
\begin{equation}
\Gamma=\frac{U(\psi_{0})U''(\psi_{0})}{U'(\psi_{0})^{2}} \le
\frac{1}{24\xi^{2}\psi_{0}^{2}}\Big(18\xi^{2}\psi_{0}^{2}+\xi -
\frac{9}{48}(1+6\xi\psi_{0}^{2}(1-6\xi))\Big).
\end{equation}

\section{Cosmology with non-minimal coupling as a dynamical system}

In this section we concentrate on investigation of the dynamics of the models 
with the quadratic potential because for other most popular forms of the potential 
functions there are 
attractors only for the negative coupling parameter. This case seems to be less 
interesting physically \cite{Tsujikawa:2004my}. For the quadratic potential
$U(\psi)=\frac{1}{2}m^{2}\psi^{2}$, $m^{2}=1$, 
the dynamical system on the invariant submanifold reduces to
\begin{align}
\varepsilon\dot{\psi} &= \frac{1}{2}y\psi^{2}(1+6\xi\psi^{2}(1-6\xi)),\nonumber \\
\varepsilon\dot{y} &= -\frac{1}{2}y\psi^{2}(1+(1-6\xi)(6\xi\psi(\psi+y)+y^{2})) - 
\frac{1}{2}\psi(1+(1-6\xi)y^{2}+6\xi(y+\psi)^{2})(2y\psi(1-3\xi)+6\xi\psi^{2}-1),
\end{align}
where a dot means differentiation with respect to $\frac{\ud}{\ud \sigma} =
\frac{1}{2}\psi^{2}(1+6\xi\psi^{2}(1-6\xi))\frac{\ud}{\ud\ln(a)}$.

\begin{figure}
\includegraphics[scale=1]{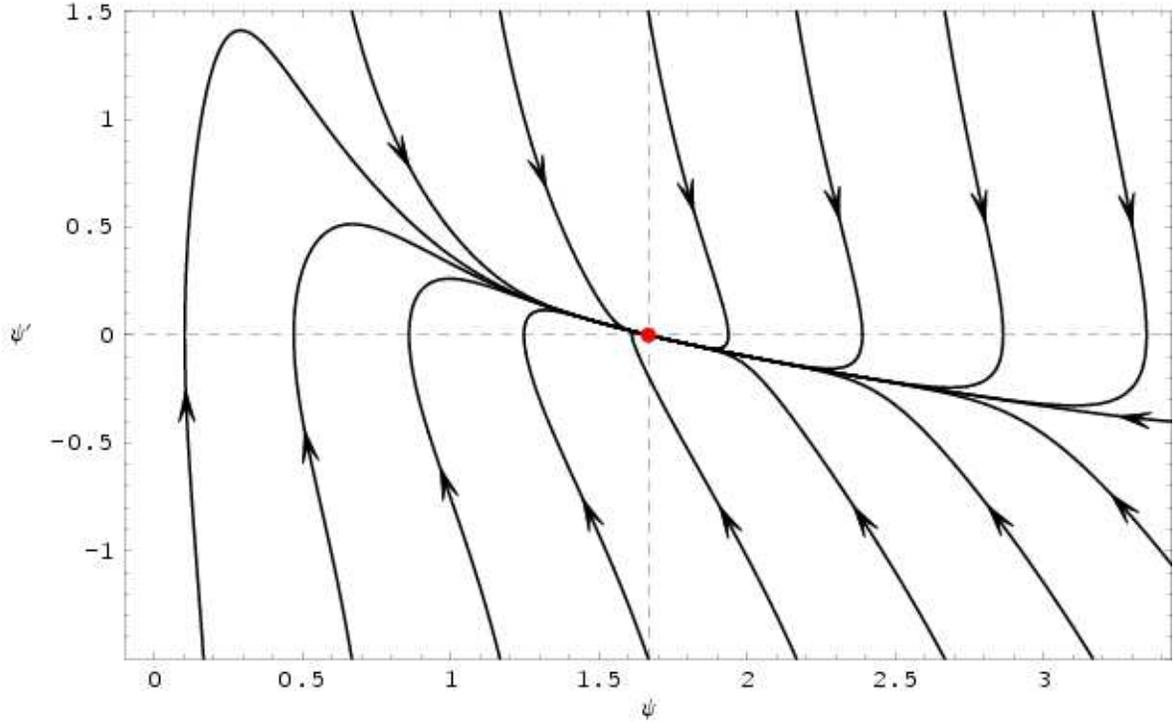}
\caption{The phase portrait for $\xi=3/50<3/25$, it represents general behavior
around the critical point of a stable node type, $w_{\text{eff}}=-1$  at this
point.}
\label{fig:1}
\end{figure}

\begin{figure}
\includegraphics[scale=1]{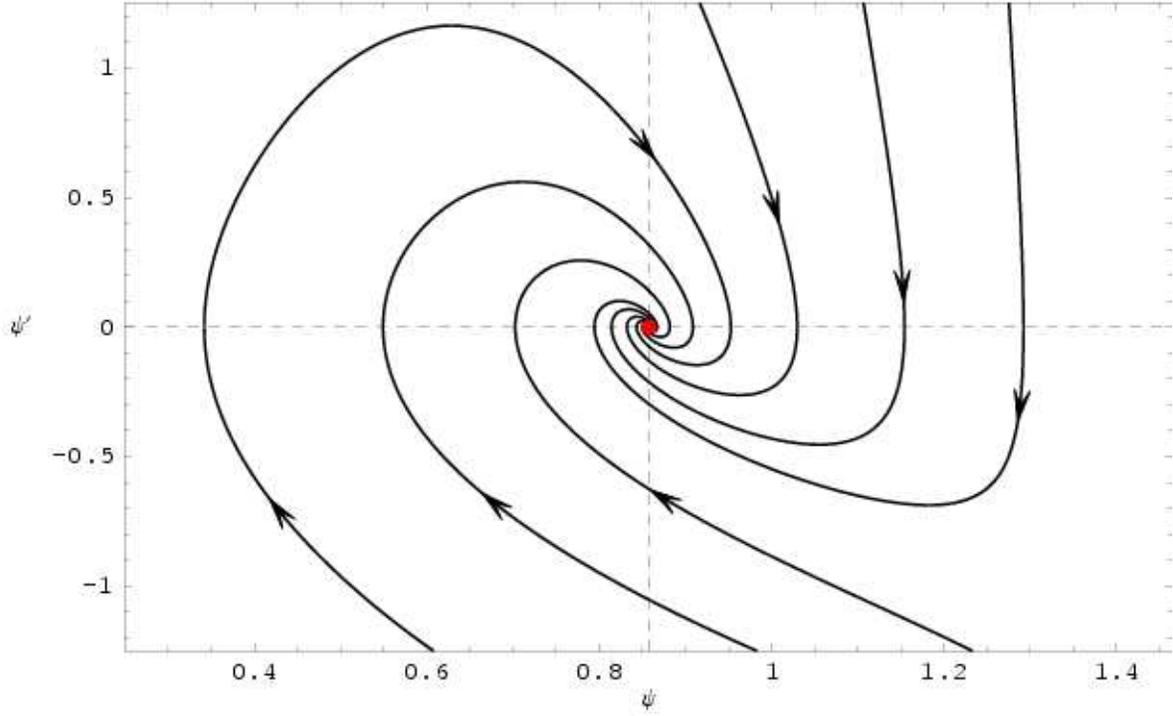}
\caption{Typical trajectories on the invariant $2$-dimensional submanifold for 
the quadratic potential function and $\xi=17/75$. The phase portrait represents 
the generic behavior around a focus type critical point. For $3/25<\xi\le1/6$ 
we have the global attractor of this type in the phase space, for $\xi>1/6$ 
the critical point of this type is no longer a global attractor.}
\label{fig:2}

\end{figure}

\begin{figure}
\includegraphics[scale=1]{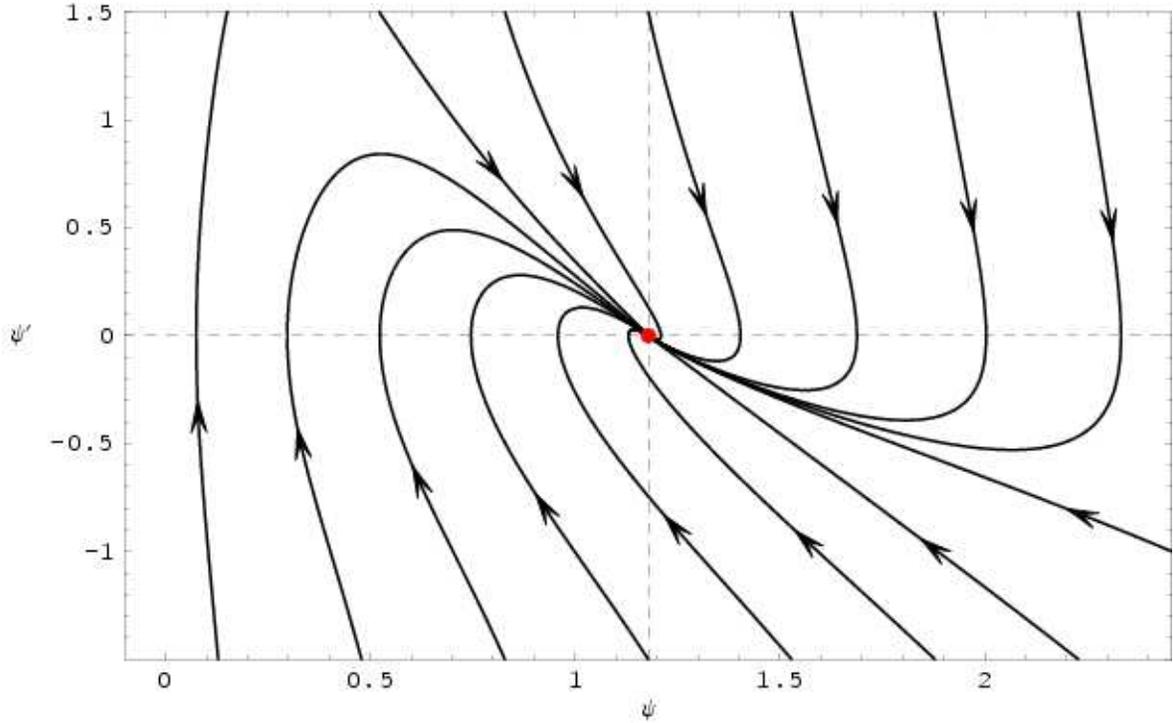}
\caption{The phase portrait for the degenerated case $\xi=3/25$.}
\label{fig:3}
\end{figure}

The linearization matrix of system (16) at the critical point $y_{0}=0$ and
$\psi_{0}^{2}=\frac{1}{6\xi}$ is
\begin{equation}
A=\left( \begin{array}{cc}
0 & \frac{1-3\xi}{6\xi} \\
-2 & -\frac{1-3\xi}{2\xi}\\
	\end{array} \right).
\end{equation}
The eigenvalues of the linearization matrix are:
$$\lambda_{1,2}=\frac{1}{4\xi}\left[ -(1-3\xi)\pm
\frac{\sqrt{3}}{3}\sqrt{\Delta}\right]$$
where $\Delta=(1-3\xi)(3-25\xi)$. The type of the critical point depends on the
value of $\Delta$.
\begin{itemize}
\item{for $0<\xi<\frac{3}{25}$:\\ $\Delta>0$ and both eigenvalues are real and negative, 
the critical point is a stable node (Fig.\ref{fig:1});}
\item{for $\frac{3}{25}<\xi<\frac{1}{3}$:\\ $\Delta<0$ we have two complex
eigenvalues and the critical point is of a focus type (Fig.\ref{fig:2});}
\item{for $\xi>\frac{1}{3}$: \\ $\Delta>0$ and $\lambda_{1}>0$, $\lambda_{2}<0$
and the critical point is of a saddle type.}
\item{for $\xi=\frac{3}{25}$ or $\xi=\frac{1}{3}$:\\ $\Delta=0$ and critical
point is degenerated (Fig.\ref{fig:3}).}
\end{itemize}

The eigenvectors are
\begin{equation}
v_{1,2}=\left[ \begin{array}{c}
1 \\
-\frac{3}{2}\pm\frac{\sqrt{3}}{2}\sqrt{\frac{3-25\xi}{1-3\xi}}\\
\end{array} \right].
\end{equation}
They are helpful in the construction of the exact solution of the linearized
system. For now, we restrict ourselves to the case of $0<\xi<3/25$, for
other cases the construction of the linearized solution is similar.
\begin{align}
\vec{x}(\sigma) &= \vec{x}(0)\exp\left\{\sigma \left( \begin{array}{cc}
0 & \frac{1-3\xi}{6\xi} \\
-2 & -\frac{1-3\xi}{2\xi}\\
        \end{array} \right)\right\}= \nonumber \\
& \left( \begin{array}{cc}
1 & 1 \\
-\frac{3}{2}+\frac{\sqrt{3}}{2}\sqrt{\frac{3-25\xi}{1-3\xi}} & 
-\frac{3}{2}-\frac{\sqrt{3}}{2}\sqrt{\frac{3-25\xi}{1-3\xi}}\\
 \end{array} \right)
 \left( \begin{array}{cc}
 \exp{\lambda_{1}\sigma} & 0 \\
 0 & \exp{\lambda_{2}\sigma} \\
 \end{array} \right) 
 \left( \begin{array}{cc}
 \frac{1}{2}+\frac{\sqrt{3}}{2}\sqrt{\frac{3-25\xi}{1-3\xi}} &
 \frac{\sqrt{3}}{3}\sqrt{\frac{3-25\xi}{1-3\xi}} \\
 \frac{1}{2}-\frac{\sqrt{3}}{2}\sqrt{\frac{3-25\xi}{1-3\xi}} &
  -\frac{\sqrt{3}}{3}\sqrt{\frac{3-25\xi}{1-3\xi}} \\
  \end{array} \right)
  \left( \begin{array}{c}
  x_{0} \\
  y_{0} \\
  \end{array} \right)
\end{align}
where $\sigma=\frac{6\xi}{1-3\xi}\ln{a}$ and $x_{0}=\psi_{i}-\psi_{0}$,
$y_{0}=\psi'_{i}-\psi'_{0}$ are initial conditions. 
The exact solutions are
\begin{align}
\psi-\psi_{0}= & \quad a^{\alpha_{1}}\left[\frac{1}{2}x_{0}+
\sqrt{3}\sqrt{\frac{1-3\xi}{3-25\xi}} 
 \left(\frac{1}{2}x_{0}+\frac{1}{3}y_{0}\right)\right] + 
a^{\alpha_{2}}\left[\frac{1}{2}x_{0}-
\sqrt{3}\sqrt{\frac{1-3\xi}{3-25\xi}}
 \left(\frac{1}{2}x_{0}+\frac{1}{3}y_{0}\right) \right], \\
\psi'-\psi'_{0} = & \quad \alpha_{1}a^{\alpha_{1}}\left[\frac{1}{2}x_{0}+
\sqrt{3}\sqrt{\frac{1-3\xi}{3-25\xi}}
 \left(\frac{1}{2}x_{0}+\frac{1}{3}y_{0}\right) \right] + 
\alpha_{2} a^{\alpha_{2}}\left[\frac{1}{2}x_{0}-
\sqrt{3}\sqrt{\frac{1-3\xi}{3-25\xi}}
  \left(\frac{1}{2}x_{0}+\frac{1}{3}y_{0}\right) \right],
\end{align}
where $x_{0}$, $y_{0}$ are initial conditions,
$\alpha_{1}=-\frac{1}{2}\Big(3-\sqrt{3}\sqrt{\frac{3-25\xi}{1-3\xi}}\Big)$ and
$\alpha_{2}=-\frac{1}{2}\Big(3+\sqrt{3}\sqrt{\frac{3-25\xi}{1-3\xi}}\Big)$.
Notice that these solutions are also valid for $\xi>1/3$ (a saddle type critical
point).

For the degenerated case $\xi=3/25$, equations (20) and (21) simplify to
\begin{align}
\psi-\psi_{0} & = x_{0} a^{-3/2}, \\
\psi'-\psi'_{0} & = -\frac{3}{2} x_{0} a^{-3/2}.
\end{align}

The same procedure can be applied to construction the linearized solution for
$\Delta<0$. Instead we can use these solutions if we notice that for
$3/25<\xi<1/3$
\begin{align}
\alpha_{1} & = -\frac{3}{2}+i\frac{\sqrt{3}}{2}\sqrt{\frac{25\xi-3}{1-3\xi}} =
-\frac{3}{2}+i\alpha, \nonumber \\
\alpha_{2} & = -\frac{3}{2}-i\frac{\sqrt{3}}{2}\sqrt{\frac{25\xi-3}{1-3\xi}} =
-\frac{3}{2}-i\alpha, \nonumber
\end{align}
then we have
\begin{align}
\psi-\psi_{0} & = a^{-3/2}\left[x_{0}\cos{(\alpha\ln{a})} +
3\alpha^{-1}\sin{(\alpha\ln{a})}\left(\frac{1}{2}x_{0}+\frac{1}{3}y_{0}\right)\right],
\\
\psi'-\psi'_{0} & = a^{-3/2}\left[y_{0}\cos{(\alpha\ln{a})} - \alpha
x_{0}\sin{(\alpha\ln{a})} - \frac{9}{2}\alpha^{-1}\sin{(\alpha\ln{a})}
\left(\frac{1}{2}x_{0}+\frac{1}{3}y_{0}\right)\right],
\end{align}
where $x_{0}$, $y_{0}$ are initial conditions, and $\alpha$ depends on the
coupling constant $\xi$ only.

The substitution of $\psi$ and $\psi'$ gives us a general formula for
$w_{\text{eff}}$ around the critical point of a given type.

On the invariant submanifold and for quadratic potential function
$U(\psi)=\frac{1}{2}\psi^{2}$ formula for equation of state parameter reduces to 
\begin{equation}
w_{X}=\frac{2}{1+6\xi\psi^{2}(1-6\xi)}\bigg\{\frac{1}{2}(1+2\xi\psi^{2}(1-6\xi))
-(1-2\xi)(1+(1-6\xi)\psi'^{2}) - 4\xi(1-3\xi)(\psi'+\psi)^{2}\bigg\}
\end{equation}

In the case when the critical point is a stable node ($0<\xi<3/25$) we have
\begin{equation}
w_{X}^{\text{mon}} = \frac{-(1-3\xi) + f_{1}(\xi,a)a^{-3/2} +
f_{2}(\xi,a)a^{-3}}
{(1-3\xi) + 6\xi(1-6\xi)\psi_{0}(A a^{\alpha_{l}}+Ba^{-\alpha_{l}})a^{-3/2} + 
3\xi(1-6\xi)(A a^{\alpha_{l}}+ B a^{-\alpha_{l}})^{2}a^{-3}},
\label{lin}
\end{equation}
where $\psi_{0}^{2}=\frac{1}{6\xi}$,
$\alpha_{l}=\frac{\sqrt{3}}{2}\sqrt{\frac{3-25\xi}{1-3\xi}}$,
\begin{align}
A & = \frac{1}{2}x_{0} +\sqrt{3}\sqrt{\frac{1-3\xi}{3-25\xi}}
\Big(\frac{1}{2}x_{0}+\frac{1}{3}y_{0}\Big), \\
B & = \frac{1}{2}x_{0} - \sqrt{3}\sqrt{\frac{1-3\xi}{3-25\xi}}
\Big(\frac{1}{2}x_{0}+\frac{1}{3}y_{0}\Big),
\end{align}
$x_{0}$ and $y_{0}$ are the initial conditions for $\psi$ and $\psi'$,
respectively, and functions $f_{1}$ and $f_{2}$ are given by
\begin{align}
f_{1}(\xi,a) & =
2\xi\psi_{0}\Big(\big(3(1-4\xi)-4\alpha_{l}(1-3\xi)\big)Aa^{\alpha_{l}} 
+ \big(3(1-4\xi)+4\alpha_{l}(1-3\xi)\big)Ba^{-\alpha_{l}}\Big), \\
f_{2}(\xi,a) & = \big(-\frac{3}{4}(3-4\xi)+15\xi(1-2\xi)\big)\Big(A a^{\alpha_{l}}+B
a^{-\alpha_{l}}\Big)^{2} + \nonumber \\
&+\alpha_{l}\big(3(1-4\xi)-8\xi(1-3\xi)\big)
\Big(A^{2}a^{2\alpha_{l}} - B^{2}a^{-2\alpha_{l}}\Big) -
\alpha_{l}^{2}(1-4\xi)\Big(Aa^{\alpha_{l}}-Ba^{-\alpha_{l}}\Big)^{2}
\end{align}
This case can be slightly simplified if we notice that there are two
independent directions leading to the critical point of the stable node type
\begin{itemize}
\item[1.]{$B=\frac{1}{2}x_{0} - \sqrt{3}\sqrt{\frac{1-3\xi}{3-25\xi}}
(\frac{1}{2}x_{0} + \frac{1}{3}y_{0})=0,$}
\item[2.]{$A=\frac{1}{2}x_{0} + \sqrt{3}\sqrt{\frac{1-3\xi}{3-25\xi}}
(\frac{1}{2}x_{0} + \frac{1}{3}y_{0})=0.$}
\end{itemize}
Then
\begin{equation}
w_{\text{mon}} = 
-\frac{(1-3\xi)+\sqrt{6\xi}(1-2\xi)x_{0}a^{\alpha_{i}}+\big(3\xi(1-2\xi) +
8\xi(1-3\xi)\alpha_{i}+(1-4\xi)\alpha_{i}^{2}\big)x_{0}^{2}a^{2\alpha_{i}}}
{(1-3\xi) + \sqrt{6\xi}(1-6\xi)x_{0}a^{\alpha_{i}} +
3\xi(1-6\xi)x_{0}^{2}a^{2\alpha_{i}}}
\end{equation}
Using general formula
\begin{equation}
\rho_{X}=\rho_{X,0}a^{-3}\exp{\Big\{-3\int_{1}^{a}\frac{w(a')}{a'}\ud a' \Big\}}
\end{equation}
we can calculate dark energy density in the case of monotonic approach to
deSitter state
\begin{equation}
\rho_{X}=\rho_{X,0} A \big(B_{0}+B_{1}a^{\alpha_{i}}+B_{2}a^{2\alpha_{i}}\big)^{\beta} 
\exp{\Big\{ -\gamma \arctan{\big(\sqrt{1-6\xi}(1+\sqrt{6\xi}x_{0}a^{\alpha_{i}})\big)}\Big\}}
\end{equation}
where
$$
A = \big(B_{0}+B_{1}+B_{2}\big)^{-\beta}
\exp{\Big\{\gamma \arctan{\big(\sqrt{1-6\xi}(1+\sqrt{6\xi}x_{0})\big)}\Big\}},
$$
$B_{0}=1-3\xi$, $B_{1}=\sqrt{6\xi}(1-6\xi)x_{0}$, $B_{2}=3\xi(1-6\xi)x_{0}^{2}$
and
$$
\beta=\frac{12\xi^{2}+8\xi(1-3\xi)\alpha_{i}+(1-4\xi)\alpha_{i}^{2}}
{2\xi(1-6\xi)\alpha_{i}},
$$
$$
\gamma=\frac{-12\xi^{2}+8\xi(1-3\xi)\alpha_{i}+(1-4\xi)\alpha_{i}^{2}}
{\xi\sqrt{1-6\xi}\alpha_{i}},
$$
$\alpha_{1}=-\frac{1}{2}\Big(3-\sqrt{3}\sqrt{\frac{3-25\xi}{1-3\xi}}\Big)$ and
$\alpha_{2}=-\frac{1}{2}\Big(3+\sqrt{3}\sqrt{\frac{3-25\xi}{1-3\xi}}\Big)$.

In the case of damped oscillatory approach to the critical point
($3/25<\xi<1/3$) and for critical point $\psi_{0}^{2}=1/6\xi$, $\psi'_{0}=0$
we have general equation of state parameter in the form
\begin{equation}
w_{X}^{\text{osc}} = \frac{-(1-3\xi)+g_{1}(\xi,a)a^{-3/2}+g_{2}(\xi,a)a^{-3}}
{(1-3\xi)+6\xi(1-6\xi)\psi_{0}h(\xi,a)a^{-3/2}
+3\xi(1-6\xi)h^{2}(\xi,a)a^{-3}
} 
\label{osc}
\end{equation}
where the functions $h$, $g_{1}$ and $g_{2}$ are
\begin{align}
h(\xi,a) & = x_{0}\cos{(\alpha_{\text{osc}}\ln{a})+\frac{3}{\alpha_{\text{osc}}}
\sin{(\alpha_{\text{osc}}\ln{a})\big(\frac{1}{2}x_{0}+\frac{1}{3}y_{0}\big)}},\\
g_{1}(\xi,a) & = 2\xi\psi_{0}\Big( (1-6\xi)h(\xi,a) - 
4(1-3\xi)\big((x_{0}+y_{0})\cos{(\alpha_{\text{osc}}\ln{a})} -
\alpha_{\text{osc}}x_{0}\sin{(\alpha_{\text{osc}}\ln{a})} - \nonumber \\
& \qquad \quad - \frac{3}{2\alpha_{\text{osc}}}\sin{(\alpha_{\text{osc}}\ln{a})
(\frac{1}{2}x_{0}+\frac{1}{3}y_{0})}\big)\Big), \\
g_{2}(\xi,a) & = \xi(1-6\xi)h^{2}(\xi,a) - \nonumber \\
& - (1-2\xi)(1-6\xi)
\Big(y_{0}\cos{(\alpha_{\text{osc}}\ln{a})} - \alpha_{\text{osc}}x_{0}\sin{(\alpha_{\text{osc}}\ln{a})} -
\frac{9}{2\alpha_{\text{osc}}}\sin{(\alpha_{\text{osc}}\ln{a})}(\frac{1}{2}x_{0}+\frac{1}{3}y_{0})\Big)^{2}
- \nonumber \\
& - 4\xi(1-3\xi)\Big((x_{0}+y_{0})\cos{(\alpha_{\text{osc}}\ln{a})} -
\alpha_{\text{osc}}x_{0}\sin{(\alpha_{\text{osc}}\ln{a})} -
\frac{3}{2\alpha_{\text{osc}}}\sin{(\alpha_{\text{osc}}\ln{a})
(\frac{1}{2}x_{0}+\frac{1}{3}y_{0})}\Big)^{2},
\end{align}
where $\alpha_{\text{osc}}=\frac{\sqrt{3}}{2}\sqrt{\frac{25\xi-3}{1-3\xi}}$ and
$x_{0}$, $y_{0}$ and $\psi_{0}$ have their usual meaning.

In the case of conformal coupling and assuming that $y_{0}=-\frac{3}{2}x_{0}$,
equation of state parameter (\ref{osc}) simplifies to
\begin{equation}
w_{X;\xi=\frac{1}{6}}^{\text{osc}} = -1 +
\frac{2}{3}a^{-3/2}\Big(\cos{(\frac{\sqrt{7}}{2}\ln{a})} +
\sqrt{7}\sin{(\frac{\sqrt{7}}{2}\ln{a})} \Big)x_{0}
-\frac{1}{6}a^{-3}\Big(4-3\cos{(\sqrt{7}\ln{a})} +
\sqrt{7}\sin{(\sqrt{7}\ln{a})}\Big)x_{0}^{2}.
\end{equation}
This expression can be easy integrated and we receive dark energy
density in the case of conformally coupled phantom scalar field
\begin{align}
\rho_{X;\xi=\frac{1}{6}}= & \rho_{X,0}\exp{\Big(-\frac{1}{48}(120-29x_{0})x_{0}\Big)}
\exp{\Big(\frac{1}{2}a^{-3/2}\big(5\cos{(\frac{\sqrt{7}}{2}\ln{a})} +
\sqrt{7}\sin{(\frac{\sqrt{7}}{2}\ln{a})}\big)x_{0}\Big)} \nonumber \\
& \exp{\Big(\frac{1}{48}a^{-3}\big(-32+3\cos{(\sqrt{7}\ln{a})} -
9\sqrt{7}\sin{(\sqrt{7}\ln{a})}\big)x_{0}^{2}\Big)},
\end{align}
where $x_{0}$ and $\rho_{X,0}$ are two constants fitted from observational data.

\begin{figure}
a)\includegraphics[scale=0.75]{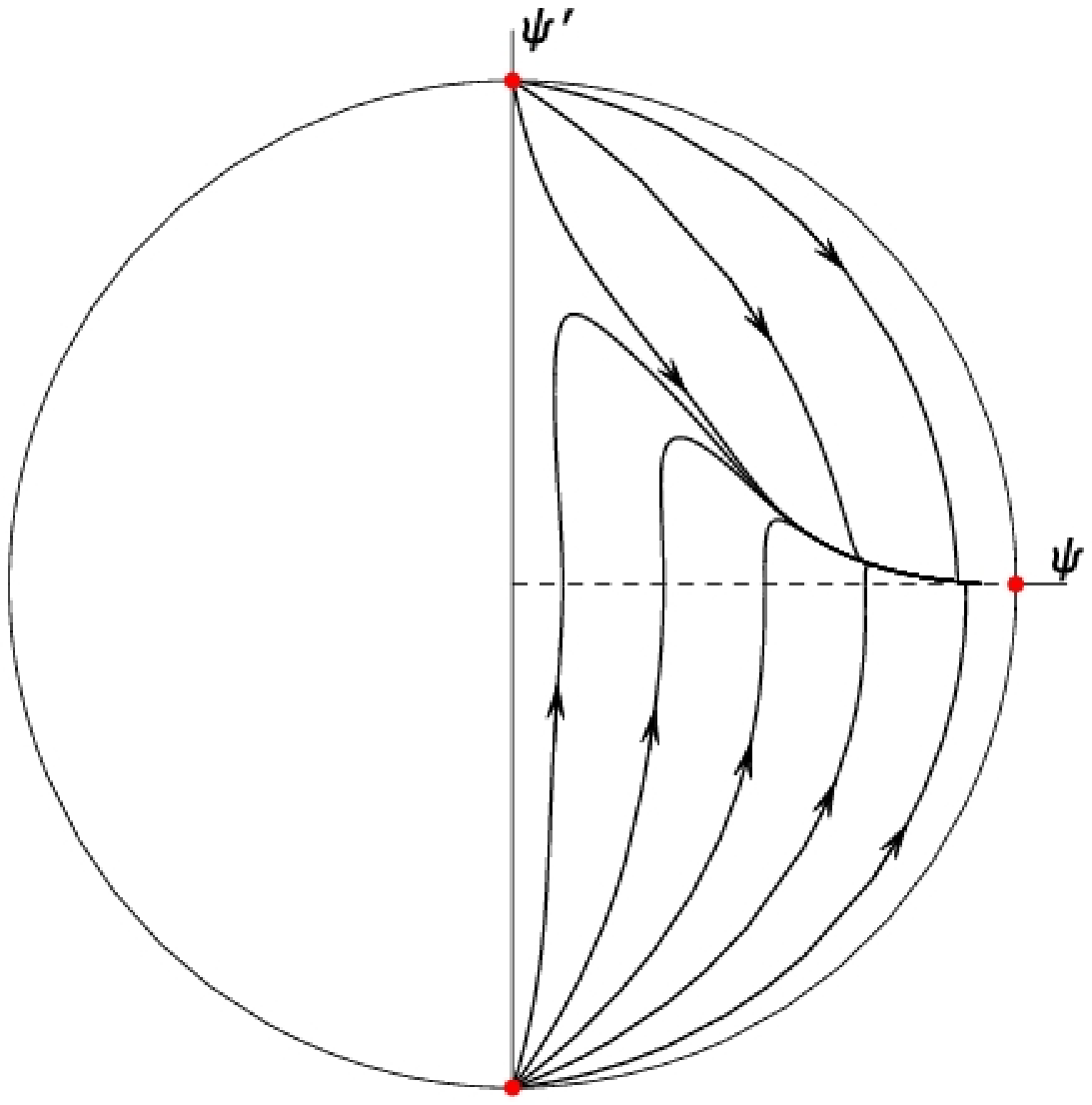}
b)\includegraphics[scale=0.75]{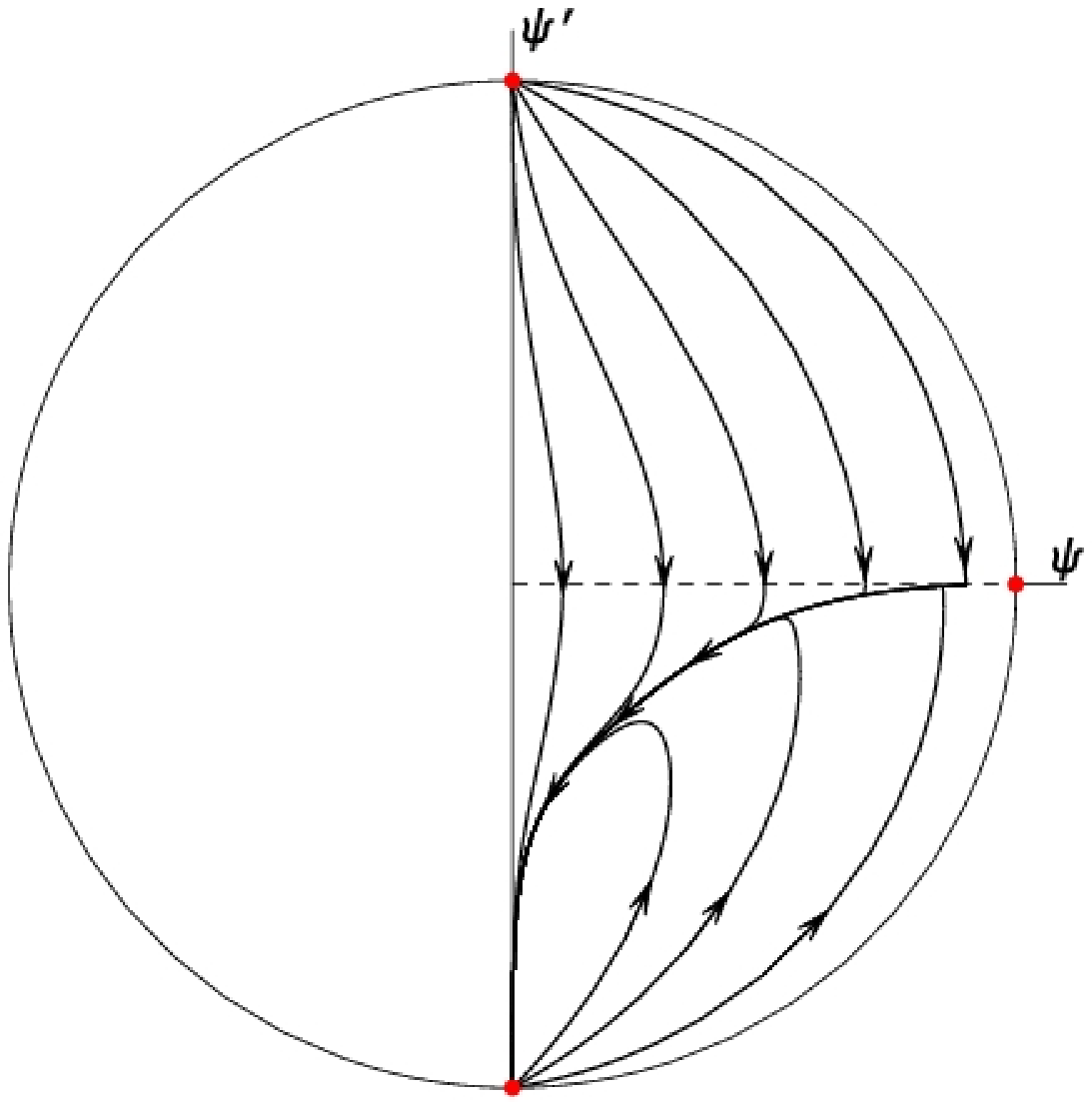}
c)\includegraphics[scale=0.75]{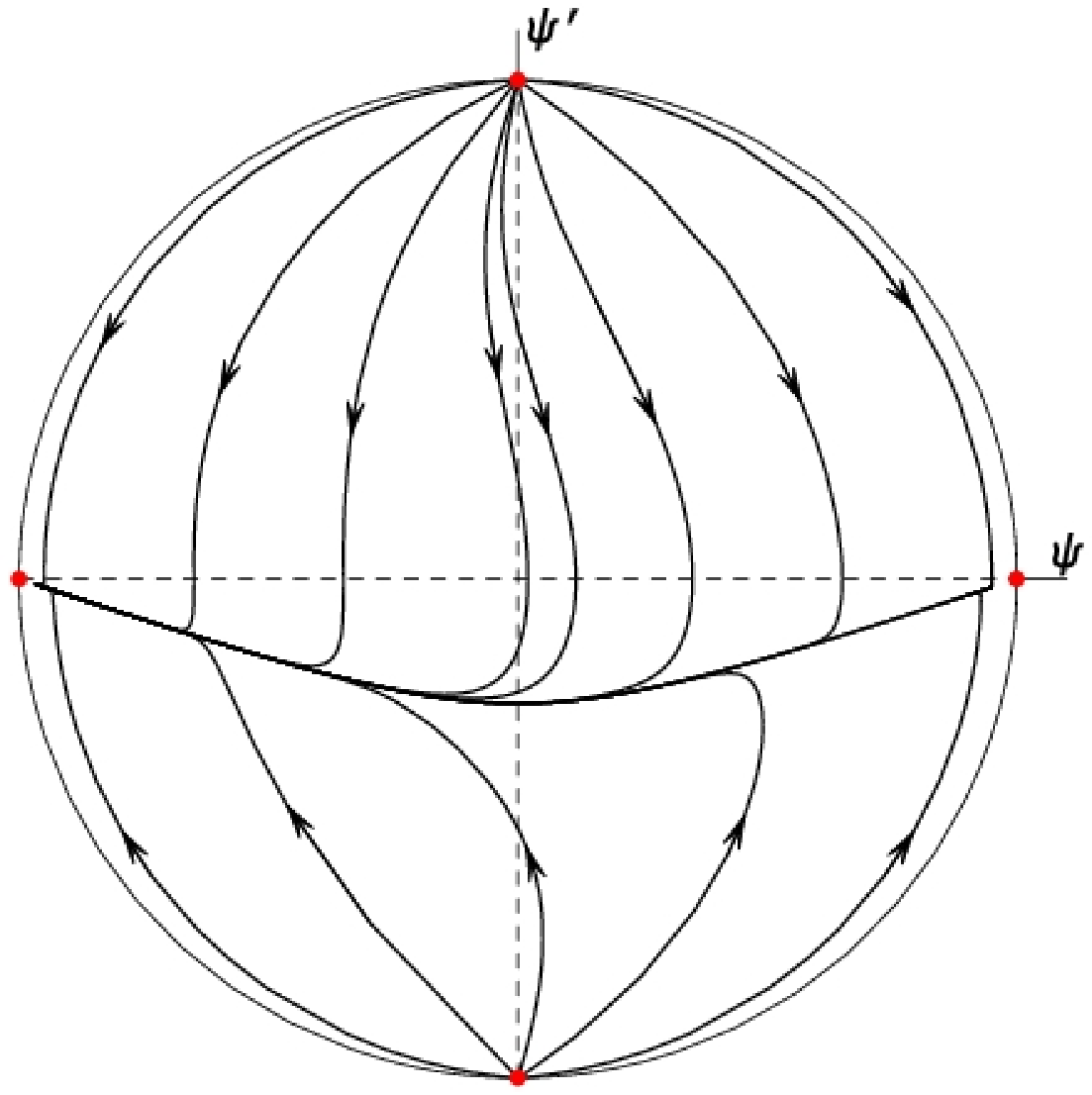}
d)\includegraphics[scale=0.75]{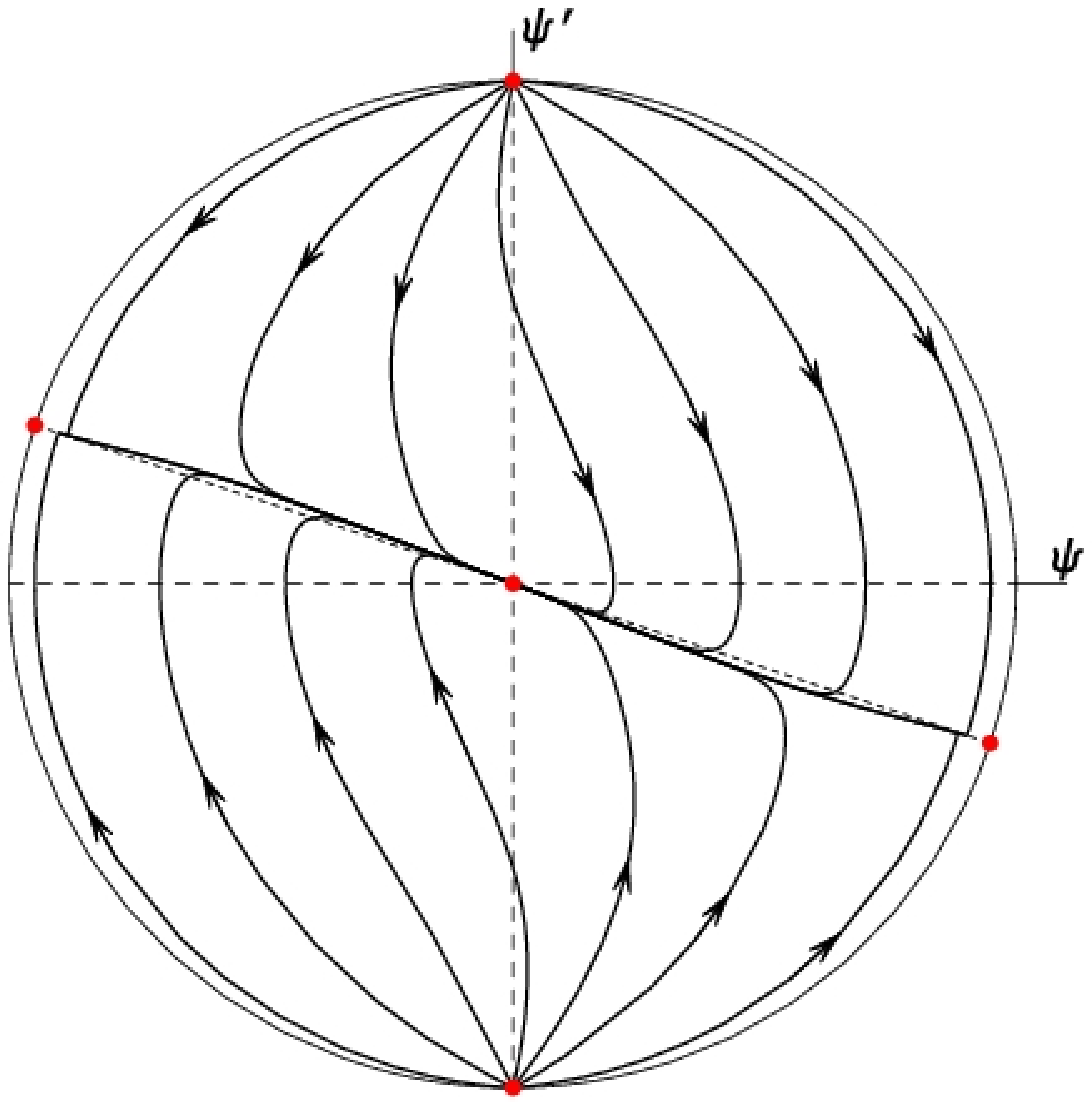}
\caption{Phase portraits with compactification at infinity for minimally coupled
phantom scalar field $\xi=0$ for various potential functions. a)
$U(\psi)\propto\psi^{2}$, critical point at $\psi'=0$ and $\psi=\infty$ plays
the
role of global attractor, $w_{X}=-1$ at this point;
b) $U(\psi)\propto\psi^{-\alpha}$ and $\alpha=1$; c) $U(\psi)\propto
\exp{(-\lambda\psi)}$ and $\lambda=1$, critical point at $\psi'=0$ and
$\psi=-\infty$ is the global attractor of the dynamical system and
$w_{X}=-1$ at this point; critical point at $\psi'=0$, $\psi=\infty$ is
a saddle type; d) $U(\psi)\propto\exp{(-\lambda\psi^{2})}$
and $\lambda=1$, $\xi=0$, the critical point at finite domain at $\psi'=\psi=0$
plays the role of global attractor and $w_{X}=-1$ at this point, the
points at infinity $\psi'\ne 0$, $\psi\ne0$ are a saddle type.}
\label{fig:4}
\end{figure}

It is interesting to consider the special case of minimal coupling $\xi=0$. The
dynamical system is in the form
\begin{equation}
\left\{ \begin{array}{l}
\displaystyle{\frac{\ud\,\psi}{\ud\,\tau} = 2 y \frac{U(\psi)}{U'(\psi)}},\\
\\
\displaystyle{\frac{\ud\,y}{\ud\,\tau} =
-(1+y^{2})\Big(6 y\frac{U(\psi)}{U'(\psi)}-1\Big)}, \\
\end{array} \right.
\end{equation}
where $\frac{\ud}{\ud\,\tau}=\frac{2}{U'(\psi)}\frac{\ud}{\ud\,\sigma}=
2\frac{U(\psi)}{U'(\psi)}\frac{\ud}{\ud\,\ln{a}}$. Above dynamical system can be
easily integrated
\begin{equation}
\ud\,y\frac{1}{1+y^{2}}= -3\ud\,\psi + \ud\,\tau,
\end{equation}
\begin{equation}
y=\psi'=\tan{(-3\psi+\Delta\,\tau + C_{0})},
\end{equation}
where
\begin{equation}
\Delta\,\tau=\frac{1}{2}\int_{i}^{a}\frac{\ud\,\ln{U(\psi)}}{\ud\,\psi}\,\ud\,\ln{a'},
\end{equation}
is the ''period of time'' between initial and final state and $C_{0}$ is the
constant of integration. The equation of state parameter assumes very simple
form for the minimal coupling
\begin{equation}
w_{X,\xi=0}=-1-2\psi'^{2}.
\end{equation}
For comparison, on the Fig.~\ref{fig:4} we present phase portraits for
minimally coupled phantom scalar field with compactification at infinity for
various potential functions. It is interesting that for every discussed form of
the potential there are always trajectories with $w_{X,\xi=0}=-1$ at a certain moment
of the evolution.

\begin{figure}
\includegraphics[scale=0.75]{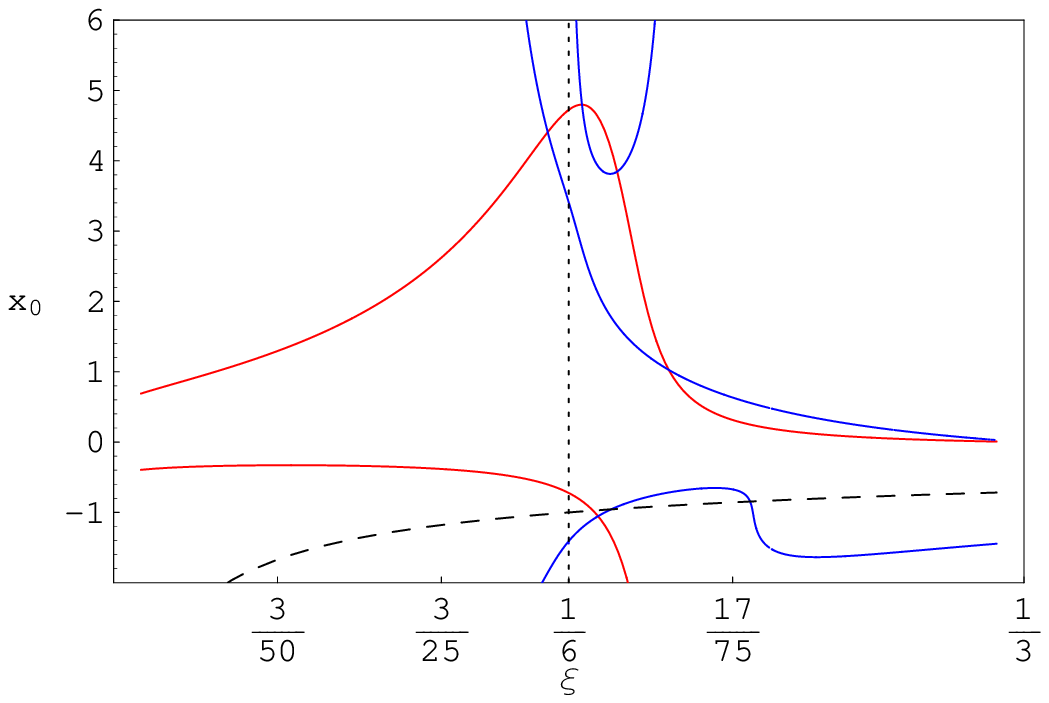}
\includegraphics[scale=0.75]{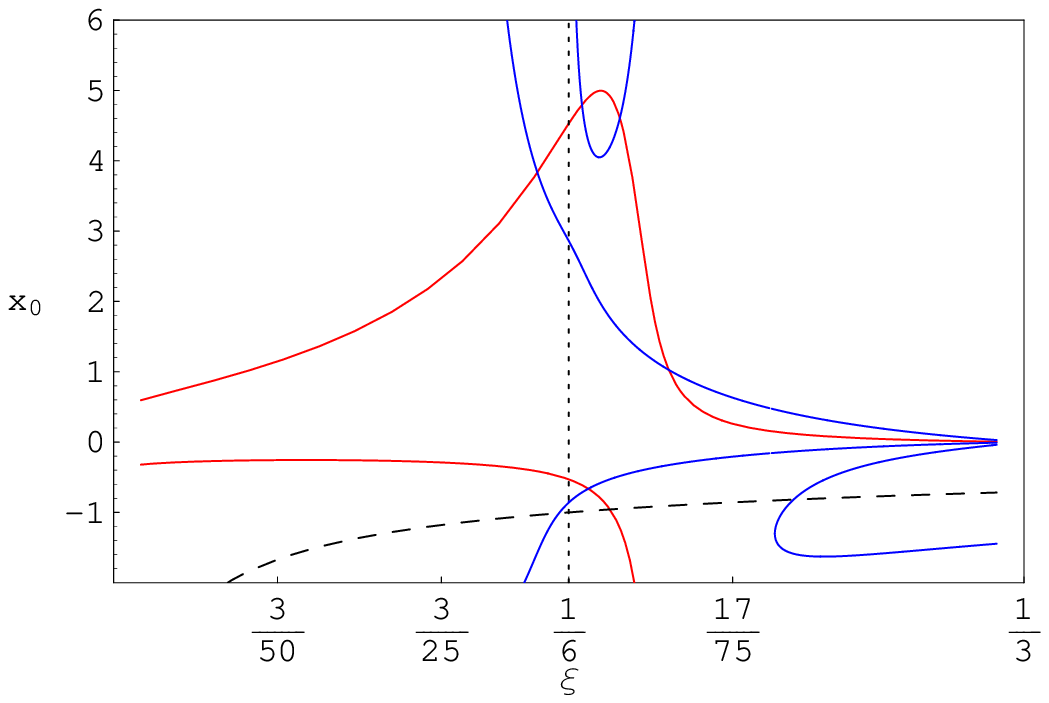}
\caption{Dependence of $x_{0}$ on the coupling constant $\xi$ for two first
terms in the Taylor series of the equation of state parameters (\ref{lin}) and
(\ref{osc}): light gray (red el.version) -- first term, dark gray (blue
el.version) -- second term and two linear parametrizations in the scale factor
(left panel) and in the redshift (right panel), values of $w_{0}$ and $w_{1}$
from \cite{Dicus:2004cp}.
The horizontal dashed line denotes unphysical limit on the values of
$x_{0}<-\frac{1}{\sqrt{6\xi}}$.
It is interesting that astronomical data together with assumption
$y_{0}=-\frac{3}{2}x_{0}$ indicate the value of $\xi$ parameter
near conformal coupling $\xi=\frac{1}{6}$.}
\label{fig:5}
\end{figure}

In current astronomical analysis of the nature of dark energy, the most popular are
linear (in the scale factor or the redshift) parametrizations of equation of
state parameter. Let us compare equations 
(\ref{lin}) and (\ref{osc}) with current estimations of coefficients of this
parametrizations. Expanding (\ref{lin}) and (\ref{osc}) in to the Taylor series
with respect to the scale factor or the redshift at the present epoch we receive
polynomial approximation of exact form of the equation of state parameter.
As an example we use two linear parametrizations and values of $w_{0}$ and $w_{1}$
estimated by Dicus and Repko \cite{Dicus:2004cp}. In the Taylor series we have three
free parameters $x_{0}$, $y_{0}$ and $\xi$, but form observations we have only
two numbers $w_{0}$ and $w_{1}$, so one free parameter should be removed. Most
convenient seems to be assumption that $y_{0}=-\frac{3}{2}x_{0}$. 
Taking in to account only two first terms
in the Taylor series we can solve system on the $\xi$ and $x_{0}$. 
On the Fig.~\ref{fig:5} we present dependence of
$x_{0}$ on the coupling parameter $\xi$ for the values of $w_{0}$ and
$w_{1}$  form \cite{Dicus:2004cp}. Intersections of this curves are solutions
on $x_{0}$ and $\xi$. It is interesting that the obtained values of the coupling
parameter $\xi$ with the assumption $y_{0}=-\frac{3}{2}x_{0}$ are close
to the conformal coupling $\xi=\frac{1}{6}$.

\section{Analysis at infinity}

The full description of a dynamical system includes also behavior of
its trajectories at infinity. It is performed in tools of the Poincare sphere
construction. In this approach we project trajectories from the center of the
unit sphere $\mathbf{S}^{2}$ onto the plane $(x,y)$ tangent to $\mathbf{S}^{2}$
either at the north or south pole (for details see \cite[p. 265]{Perko:1991}).
Due to this construction infinitely distant points of the plane are mapped into
the sphere equator, phase trajectories are mapped into corresponding curves on
the sphere but the character of critical points is conserved and new critical
points can appear at the equator. Hence the orthogonal projection of any
hemisphere onto a tangent plane gives compactified phase portrait on the plane.
In practice, polar or projective coordinates are introduced on the plane. Due to
this central projection which was introduced by Poincare we can observe how
critical points representing asymptotic states of the system are spread out
along the equator. Moreover if r.h.s. of the system are given in polynomial form
then we have simple test of structural stability of the system. In particular, a
vector field on the Poincare sphere will be structurally unstable if there are
non-hyperbolic critical points at infinity or if there is a trajectory connecting
a saddles on the equator of Poincare sphere. In opposite case if number of critical
points and limit cycles is finite dynamical system is structurally stable and
therefore generic in the Peixoto sense.

\begin{figure}
\includegraphics[scale=1]{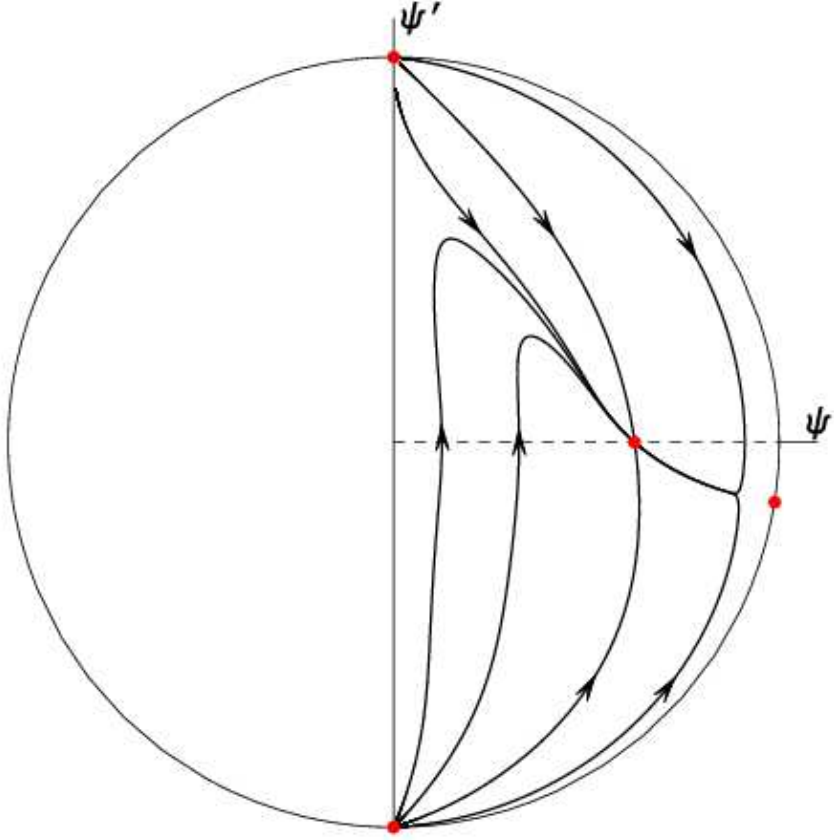}
\caption{The phase portrait with compactification at infinity for $\xi=3/50$.
The critical point at infinity is a saddle type.}
\label{fig:6}
\end{figure}

\begin{figure}
\includegraphics[scale=1]{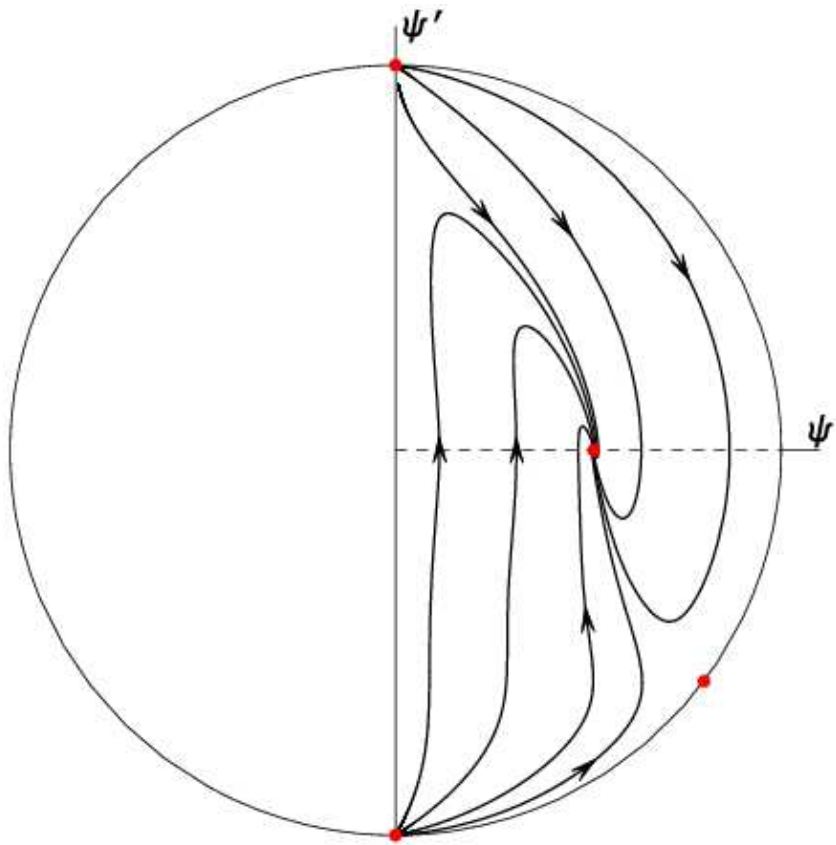}
\caption{The phase portrait with compactification at infinity for $\xi=3/20$.
The critical point of a focus type plays the role of a global attractor in the 
phase space.}
\label{fig:7}
\end{figure}

\begin{figure}
\includegraphics[scale=1]{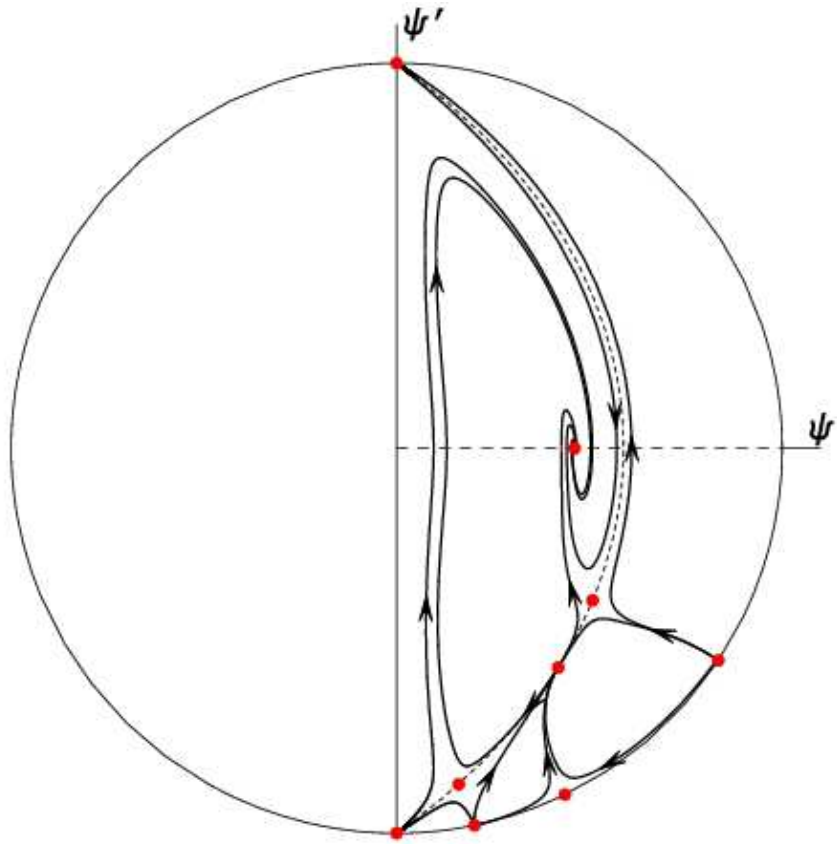}
\caption{The phase portrait with compactification at infinity for $\xi=17/75$. 
The focus type critical point is no longer a global attractor in the phase space.}
\label{fig:8}
\end{figure}

Introducing new variables $(r,\theta)$ with compactification at infinity:
\begin{align}
\psi= & \frac{r}{1-r}\cos{\theta},\nonumber \\
y = & \frac{r}{1-r}\sin{\theta}, \nonumber
\end{align}
we receive dynamical system in the form
\begin{align}
\theta'= &
-\frac{1}{2}\cos^{2}{\theta}\Big\{r^{2}\sin^{2}{\theta}
\Big((1-r)^{2}+6\xi(1-6\xi)r^{2}\cos^{2}{\theta}\Big) + \nonumber \\
 & \qquad + r^{2}\sin{\theta}\cos{\theta}\Big((1-r)^{2}+(1-6\xi)(6\xi r^{2}\cos{\theta}
(\cos{\theta}+\sin{\theta})+r^{2}\sin^{2}{\theta})\Big) + \nonumber \\
 & \qquad + \Big((1-r)^{2}+(1-6\xi)r^{2}\sin^{2}{\theta}+ 6\xi
r^{2}(\sin{\theta}+\cos{\theta})^{2}\Big) 
\Big(2r^{2}\sin{\theta}\cos{\theta}-(1-r)^{2}+6\xi
r^{2}\cos{\theta}(\cos{\theta}-\sin{\theta})\Big)\Big\},
\nonumber \\
r'= & \frac{1}{2}r(1-r)\sin{\theta}\cos{\theta}
\Big\{
r^{2}\cos^{2}{\theta}\Big((1-r)^{2}+6\xi(1-6\xi)r^{2}\cos^{2}{\theta}\Big)-
\nonumber \\
& \qquad - r^{2}\sin{\theta}\cos{\theta}\Big((1-r)^{2}+(1-6\xi)(6\xi
r^{2}\cos{\theta}
(\cos{\theta}+\sin{\theta})+r^{2}\sin^{2}{\theta})\Big) - \nonumber \\
& \qquad - \Big((1-r)^{2}+(1-6\xi)r^{2}\sin^{2}{\theta}+ 6\xi
r^{2}(\sin{\theta}+\cos{\theta})^{2}\Big) 
\Big(2r^{2}\sin{\theta}\cos{\theta}-(1-r)^{2}+6\xi
r^{2}\cos{\theta}(\cos{\theta}-\sin{\theta})\Big)
\Big\},
\nonumber
\end{align}
where a prime denotes now differentiation with respect to 
$\ud/\ud\tau = (1-r)^{4}\ud/\ud\sigma$ (Figs.~\ref{fig:6},~\ref{fig:7},~\ref{fig:8}). 

\begin{figure}
\includegraphics[scale=1]{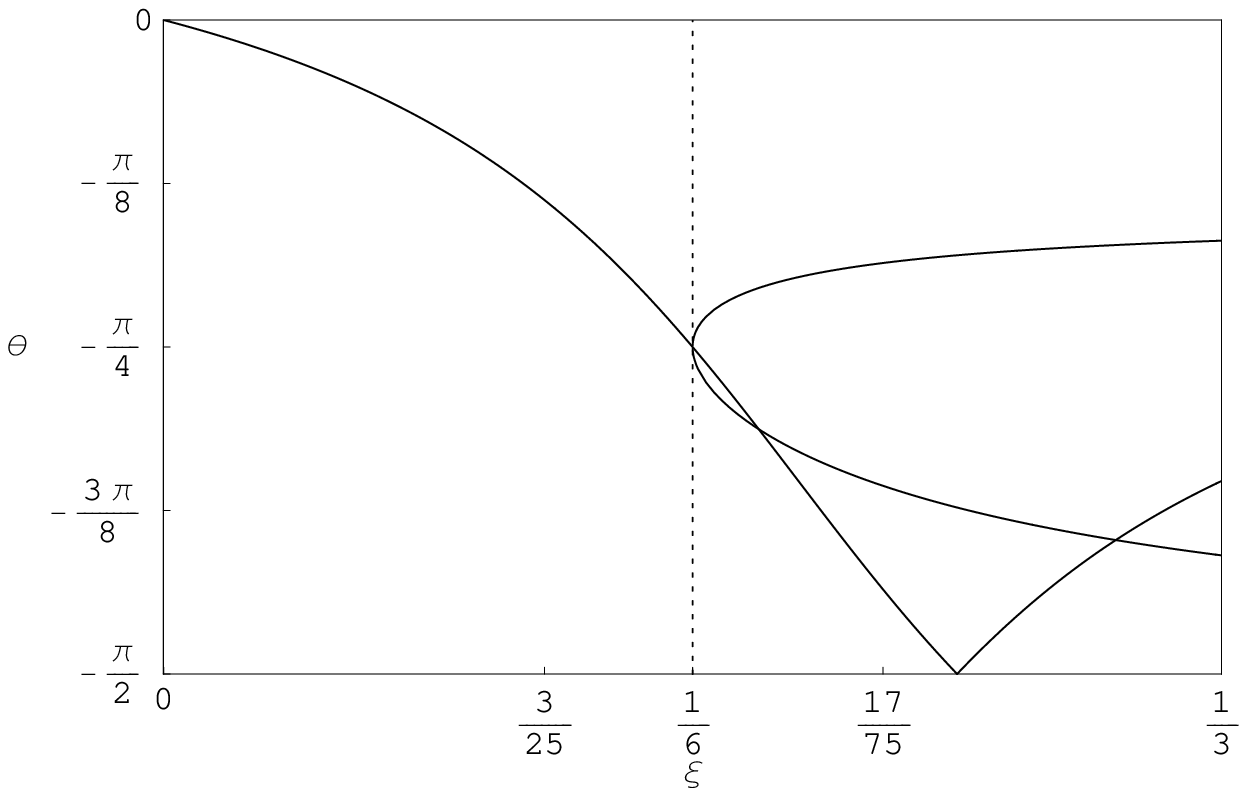}
\caption{The localization and number of critical points at infinity as a function of
coupling constant $\xi$.}
\label{fig:9}
\end{figure}

The number and character of critical points at infinity depends on the value of
$\xi$. In the general case we have critical points $(r_{0}=1,\theta_{0})$ :
$\theta=\pm\pi/2$ and solution to the equation
$$
6\xi(7-24\xi)\sin^{2}{\theta}\cos{\theta} + 18\xi\sin{\theta}\cos^{2}{\theta} +
3(1-4\xi)\sin^{3}{\theta} + 36\xi^{2}\cos^{4}{\theta}=0.
$$
This equation can be easily solved for arbitrary coupling constant $\xi$. On
Fig.~\ref{fig:9} we present location and number of critical point at infinity
as a function of coupling constant $\xi$.


\section{Conclusions}

In this paper we considered the non-minimally coupled phantom scalar field as 
the FRW model with dark energy driving the acceleration of the current universe. 
We called it the 
extended quintessence and formulated the FRW dynamics with this form of dark energy
in terms of the autonomous dynamical system. Hence we study all evolutional paths of
the extended quintessence in the phase space for all admissible initial conditions.
The structure of the phase space crucially depends on the constant of the
non-minimal coupling because right-hand sides of the dynamical system depends 
on this parameter.  We investigate bifurcations under changing of this parameter 
to distinguish some generic evolutional routes to the FRW model with the 
cosmological constant. We found that dynamical system admits the invariant 
submanifold $(\psi,\psi')$ and trajectories in the long time evolution approach 
to this submanifold of the system. The behavior of trajectories on this phase 
plane give us information how the FRW model with the cosmological constant 
appears as a final asymptotic state -- a limit set. Our main conclusion is that 
there are principally two generic evolutional scenarios of approaching the 
limit set in the future:
\begin{itemize}
\item[1.]{node type scenario,}
\item[2.]{focus type scenario (repellor-attractor scenario).}
\end{itemize}

In the first type of scenario trajectories approach the critical point monotonically without
oscillations. In the second type of scenario 
the approach to the final state is through the
characteristic infinite number of damping quasi-oscillations. 
Additionally in this case the final limit set is always a focus type and
it is achieved by evolution from an unstable node.

It is shown that the type of evolutional scenario describing the approach 
toward the stable limit set in the future crucially depends on the trace and 
the determinant of the linearization matrix of the system on the invariant 
submanifold. They can be simply expressed in the terms of slow-rolling 
parameters, i.e., in terms of geometry of a potential function of the phantom 
scalar field. It was also demonstrated how our conclusion depends on the choice 
of a special form of a potential function of the scalar field. We found that in 
the generic case two mentioned before scenarios are typical. We obtained the 
exact form of coefficient of the equation of state $w(z)$ near the present 
epoch directly from the dynamics of the model. This gave us the form of $w(z)$ 
which can be used in estimation of this parameter. Such a methodology is more 
appropriate in our belief than taking the ad-hoc form of $w(z)$ usually in a 
linear form.

\begin{acknowledgments}
This work has been supported by the Marie Curie Actions
Transfer of Knowledge project COCOS (contract MTKD-CT-2004-517186).
\end{acknowledgments}

\end{document}